%% file: main.tex
\documentclass[reprint, superscriptaddress,
amsmath,amssymb, aps, prb]{revtex4-2}

\usepackage{graphicx}
\usepackage{dcolumn}
\usepackage{bm}
\usepackage{hyperref}

\begin{document}

\preprint{APS/123-QED}

\title{Cryogenic focused-ion-beam microstructuring enabling \\quantitative $c$-axis transport measurements in Tl$_2$Ba$_2$CuO$_{6+\delta}$}

\author{Ayanesh Maiti}\email{Ayanesh.Maiti@cpfs.mpg.de}
\affiliation{Max Planck Institute for Chemical Physics of Solids, Dresden, Germany.}
\affiliation{Max Planck Institute for Structure and Dynamics of Matter, Hamburg, Germany.}
\affiliation{School of Physics and Astronomy, University of St Andrews, UK.}

\author{Carsten Putzke}\email{Carsten.Putzke@mpsd.mpg.de}
\affiliation{Max Planck Institute for Structure and Dynamics of Matter, Hamburg, Germany.}
\author{Linus Holeschovsky}
\affiliation{Max Planck Institute for Structure and Dynamics of Matter, Hamburg, Germany.}
\author{Roemer D. H. Hinlopen}
\affiliation{Max Planck Institute for Structure and Dynamics of Matter, Hamburg, Germany.}
\author{Chunyu Guo}
\affiliation{Max Planck Institute for Structure and Dynamics of Matter, Hamburg, Germany.}
\author{Dorothee Herrmann}
\affiliation{Max Planck Institute for Structure and Dynamics of Matter, Hamburg, Germany.}

\author{Seunghyun Khim}
\affiliation{Max Planck Institute for Chemical Physics of Solids, Dresden, Germany.}
\author{Berit H. Goodge}
\affiliation{Max Planck Institute for Chemical Physics of Solids, Dresden, Germany.}

\author{Andr\'e W. Tyler}
\affiliation{Interdisciplinary Research Centre in Superconductivity, University of Cambridge, UK}

\author{Michele~S.~Conroy}
\affiliation{Department of Materials, Imperial College London, UK.}
\affiliation{London Centre for Nanotechnology, Imperial College London, UK.}

\author{Andreas W. Rost}
\affiliation{School of Physics and Astronomy, University of St Andrews, UK.}

\author{Andrew P. Mackenzie}\email{Andrew.Mackenzie@cpfs.mpg.de}
\affiliation{Max Planck Institute for Chemical Physics of Solids, Dresden, Germany.}
\affiliation{School of Physics and Astronomy, University of St Andrews, UK.}
\affiliation{Interdisciplinary Research Centre in Superconductivity, University of Cambridge, UK}

\author{Philip J. W. Moll}\email{Philip.Moll@mpsd.mpg.de}
\affiliation{Max Planck Institute for Structure and Dynamics of Matter, Hamburg, Germany.}

\date{\today}

\begin{abstract}
Absolute transport measurements in correlated quantum materials are often limited by disorder, inhomogeneity, geometric uncertainty, and small crystal size. Focused ion beam (FIB) technology offers a route to overcome many of these limitations by enabling transport devices with precisely defined geometry to be fabricated from lamellae extracted from carefully selected regions of a crystal, but its application to cuprate superconductors has been hindered by ion-beam-induced damage. Here we study the clean overdoped cuprate Tl2201 and show that conventional FIB processing causes thermally driven oxygen loss, while cryogenic FIB microstructuring largely suppresses this degradation and preserves the crystal structure from the bulk to the atomic scale. Microstructured devices quantitatively reproduce established in-plane resistivity and Hall carrier density measurements without rescaling. Applying this approach to $c$-axis transport, we obtain absolute $\rho_c(T)$ values approximately three times larger than previously reported, bringing the transport anisotropy into quantitative agreement with the known Fermi surface geometry within an isotropic relaxation-time approximation. These results resolve a long-standing discrepancy between transport and quantum oscillation measurements in overdoped Tl2201 and establish cryogenic FIB microstructuring as a route to reliable quantitative transport measurements in quantum materials where disorder, inhomogeneity, geometry, or small crystal size have previously limited experimental accuracy.
\end{abstract}

\maketitle

\input{Tex/intro}
\input{Tex/results}
\input{Tex/discussion}
\input{Tex/methods}

\section*{Acknowledgements}
Research in Dresden benefits from the environment provided by the DFG Cluster of Excellence ctd.qmat (EXC2147, Project ID 390858490). AM acknowledges funding from the Max Planck Society via the MPGC-QM and IMPRS-CPQM programmes. RDHH acknowledges sponsorship provided by the Alexander von Humboldt Foundation. AWR acknowledges support from the Engineering and Physical Sciences Research Council (grant EP/P024564/1). This work was funded by the Deutsche Forschungsgemeinschaft (DFG, Project ID 501654252).

\appendix
\setcounter{figure}{0}
\renewcommand{\thefigure}{A\arabic{figure}}

\bibliographystyle{apsrev4-2}
\bibliography{Tex/references}

\input{Tex/supplement}

\end{document}

%% file: Tex/intro.tex
\begin{figure*}[t]
\centering
\includegraphics[scale=0.82]{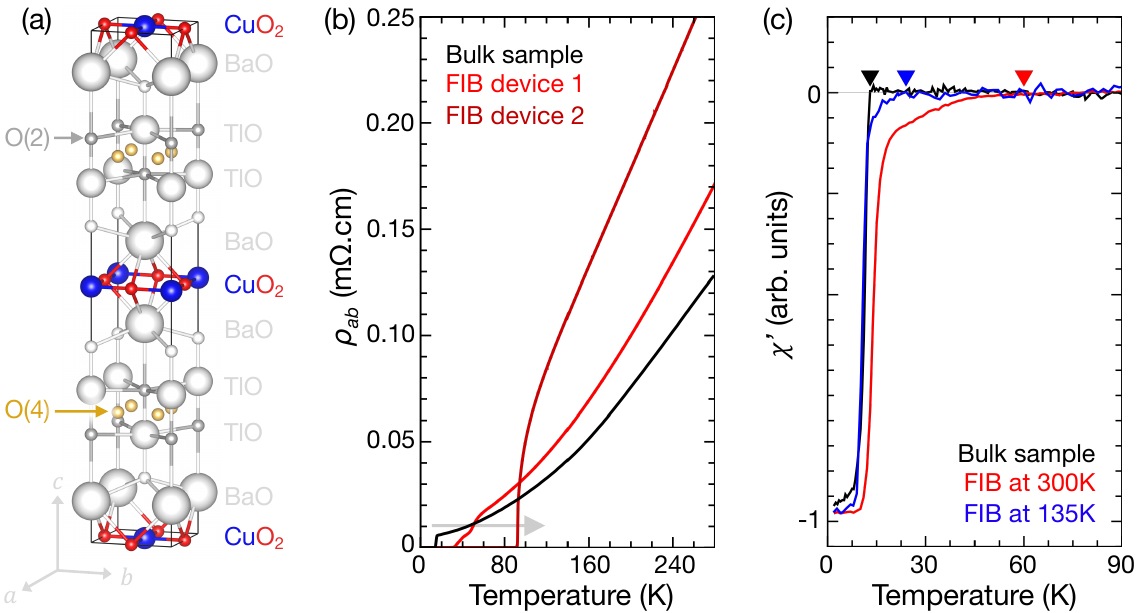}
\caption{
{\bf Cryogenic suppression of FIB-induced damage in Tl2201.}
{\bf a.} Crystal structure of Tl$_2$Ba$_2$CuO$_{6+\delta}$, highlighting the interstitial O(4) sites and the lattice O(2) sites that control hole doping and are depleted by vacuum annealing.
{\bf b.} Transport measurements on a Tl2201 single crystal before and after device fabrication using standard FIB microstructuring techniques. Ion-beam exposure increases $T_{\rm c}$ in overdoped Tl2201 (grey arrow), consistent with oxygen loss driving the material towards optimal doping. The FIB-patterned devices shown here are likely inhomogeneous and do not represent the intrinsic behaviour of Tl2201 at any well-defined doping level.
{\bf c.} Magnetic susceptibility measurements on a pristine Tl2201 crystal (black), compared with sections extracted by FIB milling at 300\,K (red) and 135\,K (blue). Room-temperature processing produces a substantial shift and broadening of the superconducting transition, indicating changes in oxygen content and increased spatial inhomogeneity. Both effects are strongly suppressed when milling is performed at cryogenic temperature, demonstrating that cryo-FIB processing largely suppresses FIB-induced changes to the doping state and electronic homogeneity.
}
\label{fig1}
\end{figure*}

\begin{figure*}[t]
\centering
\includegraphics[scale=0.82]{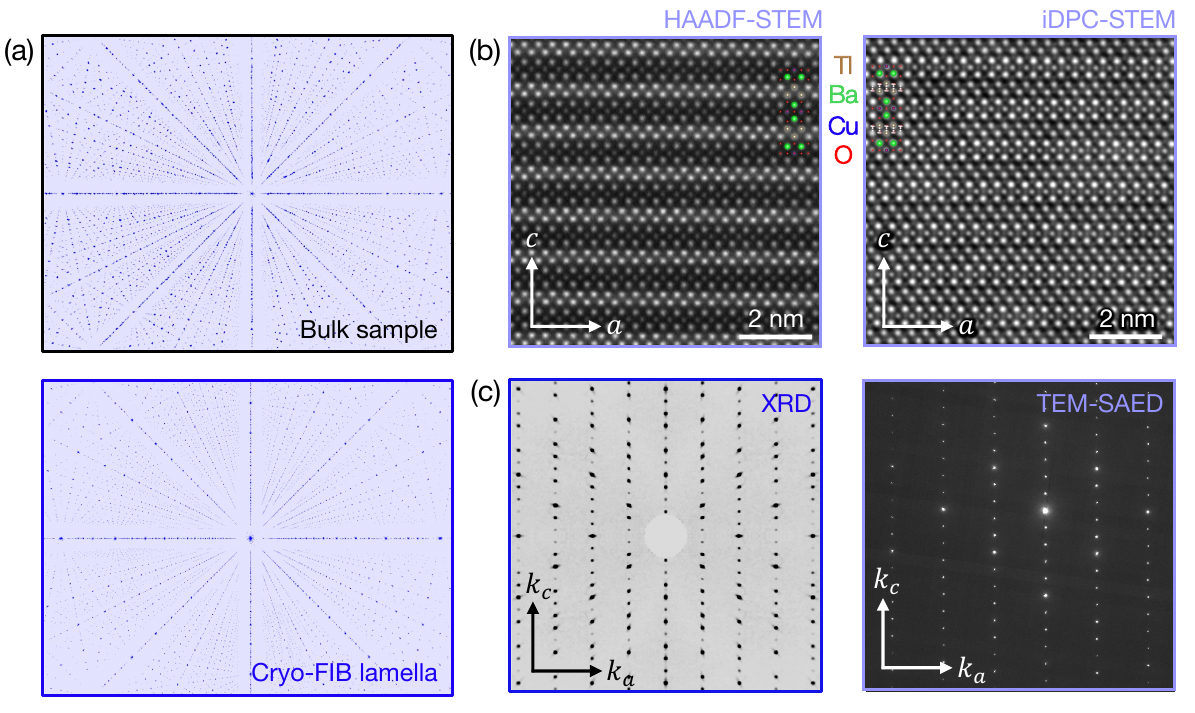}
\caption{
{\bf Structural integrity of cryo-FIB microstructured Tl2201.}
{\bf a.} Gnomonic projections of single-crystal XRD datasets acquired from a bulk Tl2201 crystal and a cryo-FIB-milled lamella. Both datasets exhibit identical Bragg reflections consistent with the known crystal structure of Tl2201, indicating that cryogenic microstructuring preserves long-range crystalline order.
{\bf b.} Atomic-resolution images of a cryo-FIB lamella acquired using HAADF-STEM and iDPC imaging. The observed atomic arrangement is consistent with the expected Tl2201 crystal structure, demonstrating preservation of the lattice at the atomic scale.
{\bf c.} Single-crystal XRD precession images compared with selected area electron diffraction (SAED) patterns acquired from cryo-FIB lamellae. Equivalent diffraction features are observed in both measurements, with no evidence of secondary phases or additional diffraction peaks introduced during cryogenic microstructuring.
}
\label{fig2}
\end{figure*}

\section*{Introduction}

Electrical transport is one of the most direct probes of the low-energy electronic structure of quantum materials, providing access to scattering processes, Fermi surface geometry, and quasiparticle coherence. In high-$T_{\rm c}$ cuprate superconductors, transport studies have been central to mapping the phase diagram, including the identification of pseudogap, strange metal, and Fermi liquid regimes in the normal state~\cite{Keimer2015}. In practice, however, obtaining quantitative transport measurements is not straightforward: surface chemistry can limit the formation of reliable electrical contacts~\cite{Ekin2006}, while disorder and inhomogeneity can obscure the intrinsic normal-state response~\cite{Alloul2024}.

\enlargethispage{\baselineskip}
In principle, focused ion beam (FIB) microstructuring offers a powerful route to overcoming many of these experimental limitations~\cite{Moll2018}. It allows sub-micron control over device geometry and current paths, while enabling transport measurements on smaller, more homogeneous regions of a crystal that are less likely to contain extended defects. The reduced device cross-section also increases the measured resistance by several orders of magnitude, improving the signal-to-noise ratio and the precision with which the absolute resistivity can be determined.

Small, precisely controlled microstructured devices are ideally fabricated from fully FIB-sculpted lamellae cut from bulk single crystals~\cite{Moll2018,Bachmann2019}. This approach has been highly successful across a range of quantum materials, particularly in chemically robust systems~\cite{Kushwaha2017,Putzke2020}, but it is much more difficult for reactive or heat-sensitive materials. This is especially true for cuprate superconductors, where prolonged FIB exposure can produce spatially inhomogeneous oxygen stoichiometry and, consequently, a distribution of local superconducting transition temperatures. Extending the lamella-based microstructuring approach for precision transport measurements in cuprates therefore requires a fabrication protocol that preserves the intrinsic electronic state of the material.

Tl$_2$Ba$_2$CuO$_{6+\delta}$ (Tl2201) in the far-overdoped regime is an ideal system on which to develop such an approach. Among the cuprates, Tl2201 possesses one of the longest electronic mean free paths (exceeding 500\,\AA\ in far overdoped samples)~\cite{Vignolle2008,Rourke2010,Bangura2010}. Its normal state displays many characteristics of a conventional Fermi liquid and has been characterised across several experimental probes~\cite{Wade1994,Carrington1996,Mackenzie1993,Mackenzie1996,Tyler1998,Hussey1996,Hussey2003,Plate2005,Abdel-Jawad2006,Analytis2007,Vignolle2008,Rourke2010,Bangura2010}. This existing characterisation provides a stringent benchmark for in-plane transport: any deviation of a microstructured device from established bulk transport and Hall coefficients can be attributed directly to processing-induced damage, rather than to intrinsic sample variability. Furthermore, Tl2201 is highly reactive and sensitive to heating and oxygen loss, making it necessary to anneal samples immediately prior to all experiments~\cite{Tyler1997}. This sensitivity makes Tl2201 a particularly stringent test of a lamella-based fabrication protocol.

\enlargethispage{\baselineskip}

Here we show that cryogenic FIB processing suppresses ion-beam-induced damage in Tl2201, yielding microstructured devices that reproduce bulk in-plane transport and Hall coefficients while preserving the crystal structure down to the atomic scale. Combined with a surface preparation and metallisation procedure that yields reproducible ohmic contacts, this approach enables quantitative transport measurements on microstructured devices. Extending this approach to $c$-axis transport, we obtain a resistive anisotropy in quantitative agreement with expectations from the known Fermi surface warping, resolving a long-standing puzzle in Tl2201. More broadly, these results establish cryogenic FIB microstructuring as a route to quantitative transport measurements in correlated quantum materials where crystal size, geometry, transport anisotropy, or beam damage have previously precluded reliable absolute determination.

\enlargethispage{\baselineskip}

%% file: Tex/results.tex
\section*{Results}
\subsection*{Cryogenic FIB suppresses oxygen loss}

The hole doping of Tl$_2$Ba$_2$CuO$_{6+\delta}$ is controlled by the oxygen content at the interstitial O(4) site indicated in Fig.~\ref{fig1}a~\cite{Wagner1997,Peets2008,Shimakawa1993}. This makes the electronic properties of Tl2201 highly sensitive to oxygen loss during processing. Fig.~\ref{fig1}b shows $\rho_{ab}(T)$ for FIB microstructured devices, fabricated from an annealed single crystal, following the procedure of Moll~\textit{et al.}~\cite{Moll2018,Bachmann2019}. In contrast to previous observations in YBCO and LSCO, where ion-beam exposure increases resistivity and suppresses $T_{\rm c}$~\cite{Sefrioui2001,Caruso2023,Hensel1997}, the FIB-sculpted Tl2201 devices instead show an enhancement of $T_{\rm c}$. This behaviour cannot be explained by disorder-induced suppression of superconductivity~\cite{Rullier-Albenque2001}, and instead indicates that FIB processing drives the overdoped material towards optimal doping, likely through a reduction in oxygen content.

To further probe this oxygen loss, Fig.~\ref{fig1}c shows magnetic susceptibility measurements (see Methods) on a pristine Tl2201 crystal compared with lamellae extracted by FIB milling at 300~K and 135~K. Room-temperature FIB processing produces a substantial broadening and shift of the superconducting transition, indicating both a change in oxygen content and increased spatial inhomogeneity in the doping state. This inhomogeneity is consistent with a spatially uneven loss of interstitial oxygen, likely driven by FIB-induced heating~\cite{Kim2011} in the high-vacuum FIB environment, which acts as a rapid, spatially localised vacuum anneal.

Under sufficiently severe conditions, oxygen loss is known to extend beyond the interstitial O(4) sites to the lattice O(2) sites (Fig.~\ref{fig1}a)~\cite{Wagner1997}. This can result in structural degradation and irreversible damage of the microstructures. Cryogenic FIB tools have previously been shown to mitigate beam-induced heating~\cite{Kim2011,Narayan2015,Noble2024}. Figure~\ref{fig1}c shows that both the shift and broadening of the transition in FIB-patterned Tl2201 are strongly suppressed when milling is performed at cryogenic temperature, demonstrating that cryo-FIB considerably improves the device oxygen stoichiometry and electronic homogeneity relative to standard room-temperature processing. Importantly, the small remaining inhomogeneity is reversible and can be removed by low-temperature annealing of the device, as described in Methods below.

\subsection*{Cryo-FIB preserves the crystal structure of Tl2201}

To determine whether cryogenic FIB processing preserves the crystal structure of Tl2201, we characterised cryo-FIB lamellae using single-crystal X-ray diffraction (XRD) and scanning transmission electron microscopy (STEM), as described in Methods.

Figure~\ref{fig2}a shows the gnomonic projections of a single-crystal XRD measurement on a cryo-FIB lamella mounted on a TEM grid and measured in transmission geometry, compared with an equivalent dataset from a bulk Tl2201 crystal. Fewer Bragg reflections are observed in the lamella dataset due to geometric constraints imposed by the sample mounting. Nonetheless, both datasets exhibit identical Bragg reflections consistent with the tetragonal crystal structure of Tl2201~\cite{Shimakawa1993}, demonstrating that cryogenic microstructuring preserves long-range crystalline order.

\begin{figure*}[t]
\centering
\includegraphics[scale=0.82]{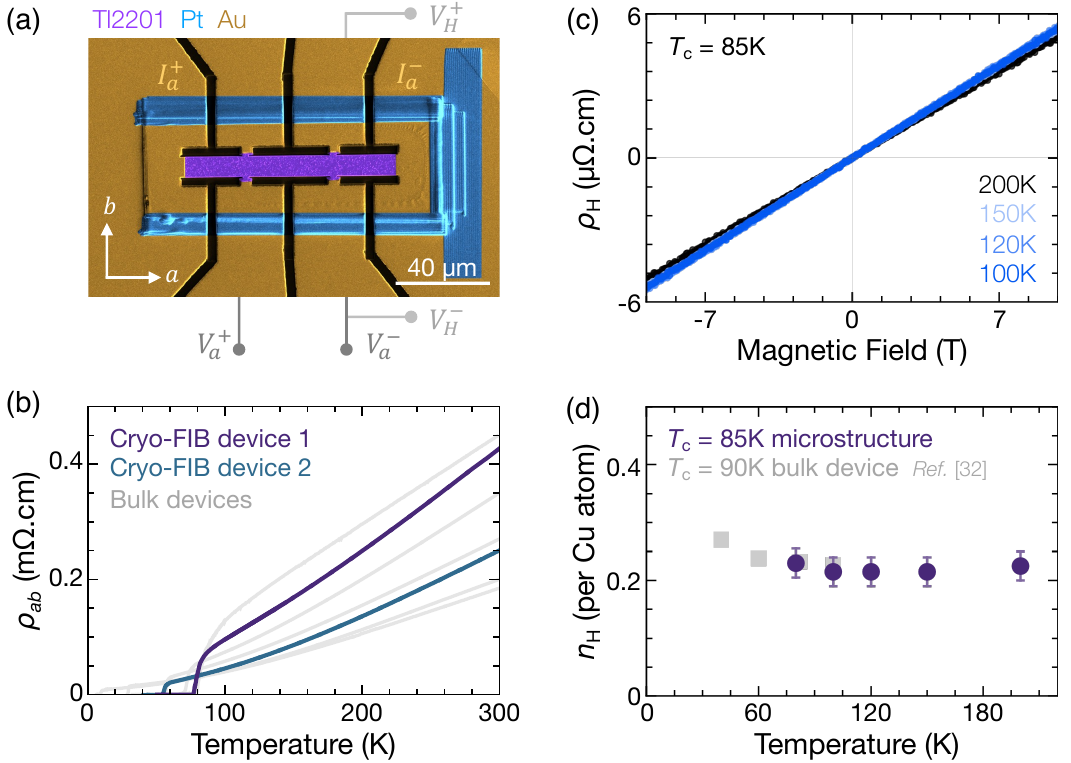}
\caption{
{\bf Validation of transport measurements in cryo-FIB microstructured Tl2201.}
{\bf a.} False-coloured SEM image of a representative $ab$-plane transport device with a six-terminal geometry, enabling simultaneous four-point measurements of the in-plane resistivity and Hall effect.
{\bf b.} In-plane resistivity of annealed cryo-FIB microstructured devices. The measured temperature dependence is quantitatively consistent with transport measurements on bulk single crystals, indicating that cryo-FIB processing preserves the intrinsic in-plane transport properties of Tl2201.
{\bf c.} Antisymmetrised Hall resistivity of a cryo-FIB device near optimal doping above $T_{\rm c}$. The Hall response is linear in magnetic field, allowing reliable extraction of the Hall coefficient. {\bf d.} The corresponding Hall carrier density $n_{\rm H}$ is consistent with previous measurements on bulk single crystals~\cite{Putzke2021}, further validating the accuracy of transport measurements performed on cryo-FIB microstructures.}
\label{fig3}
\end{figure*}

\begin{figure*}[t]
\centering
\includegraphics[scale=0.82]{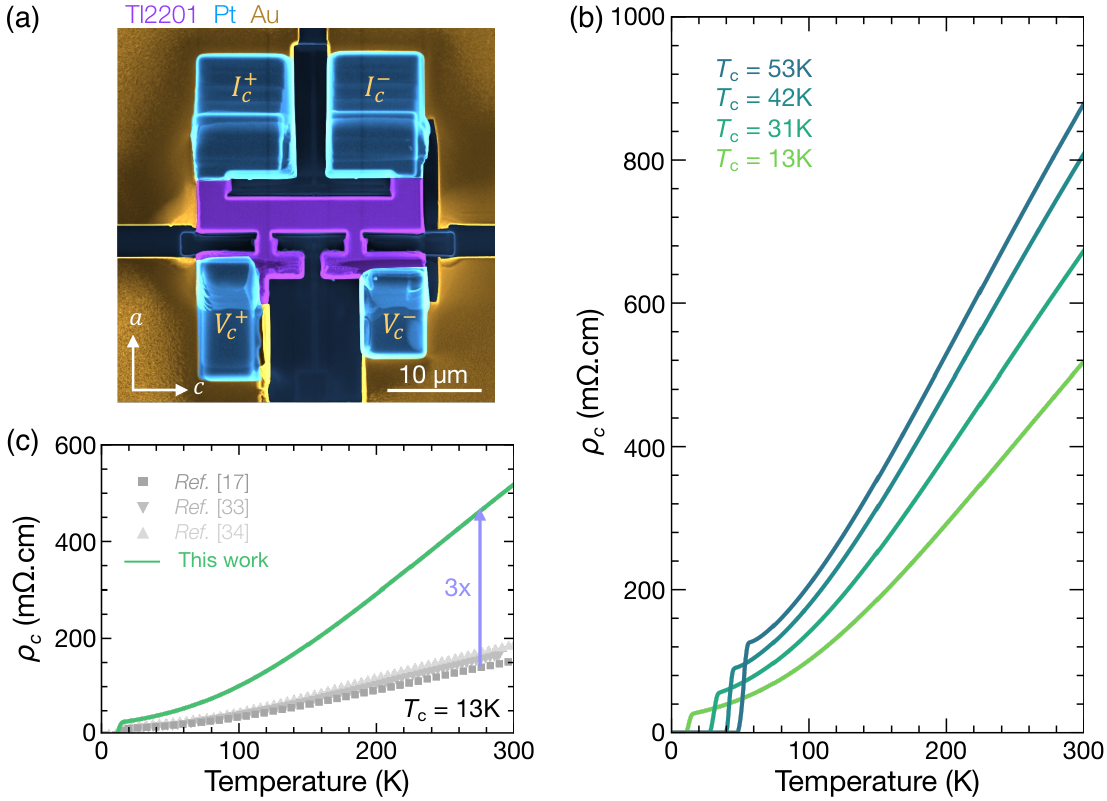}
\caption{
{\bf Intrinsic $c$-axis transport in cryo-FIB microstructured Tl2201.}
{\bf a.} False-coloured SEM image of a representative $c$-axis transport device used for four-point resistivity measurements. The microstructured geometry enables precise determination of the current path and sample dimensions, allowing accurate measurement of the absolute $c$-axis resistivity.
{\bf b.} Temperature dependence of the $c$-axis resistivity $\rho_c$ measured in cryo-FIB microstructured Tl2201 devices spanning a range of hole dopings. The measured behaviour is qualitatively consistent with previous reports on bulk single crystals, exhibiting metallic transport at high temperature and a systematic reduction in $\rho_c$ with increasing hole doping.
{\bf c.} Comparison of $\rho_c(T)$ obtained from a $T_{\rm c}=13$~K cryo-FIB microstructure against previous reports on bulk single crystals~\cite{Hussey2003,Ma2007,Manako1992}. While the temperature dependence remains largely unchanged, the absolute values of $\rho_c$ are systematically larger by a factor of approximately $2.9\pm0.3$ over the entire temperature range.}
\label{fig4}
\end{figure*}

\begin{figure*}[t]
\centering
\includegraphics[scale=0.82]{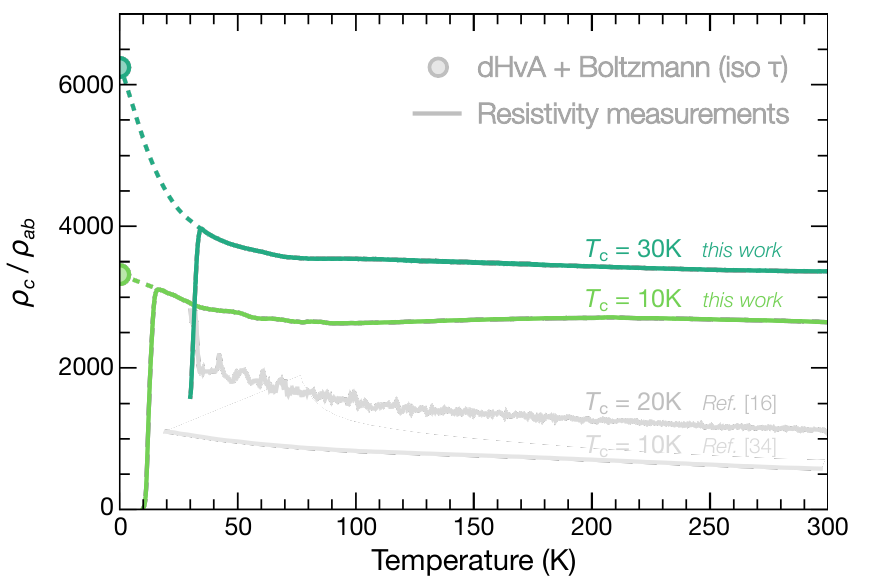}
\caption{{\bf Intrinsic transport anisotropy in Tl2201.}
Transport anisotropy $\rho_c/\rho_{ab}$ obtained from the cryo-FIB $c$-axis resistivity measurements of Fig.\ref{fig4} and bulk in-plane transport data of Fig.\ref{fig3}. The resulting anisotropy is systematically larger than previous transport estimates~\cite{Manako1992,Hussey1996} by approximately the same factor by which the absolute $c$-axis resistivity is enhanced.
For comparison, zero-temperature anisotropy values calculated from the known dHvA and AMRO Fermi surface geometry of overdoped Tl2201, assuming an isotropic relaxation time~\cite{Rourke2010}, are also shown. Dashed lines are a guide to the eye, indicating a possible smooth extrapolation of the electrical transport anisotropy below $T_{\rm c}$. In contrast to previous transport measurements, the anisotropy obtained from cryo-FIB microstructures is in quantitative agreement with the quantum-oscillation estimates, indicating that a strongly anisotropic scattering rate is not required to reconcile transport and Fermi-surface probes.
}
\label{fig5}
\end{figure*}

Figure~\ref{fig2}b shows atomic-resolution STEM images of a cryo-FIB lamella prepared from a pre-annealed single crystal at $210$\,K, acquired using high-angle annular dark field (HAADF) and integrated differential phase contrast (iDPC) imaging techniques. The observed atomic arrangement is in excellent agreement with the crystal structure of Tl2201, demonstrating preservation of the lattice at the atomic scale even after cryo-FIB thinning to less than 50~nm. Oxygen-sensitive iDPC imaging reveals no detectable signatures of oxygen disorder. Consistent with this, Fig.~\ref{fig2}c shows that the electron diffraction patterns acquired from cryo-FIB lamellae are equivalent to single-crystal XRD precession images, with no evidence of secondary phases or superlattice reflections associated with oxygen inhomogeneity.

\enlargethispage{\baselineskip}
These observations indicate that cryo-FIB processing preserves the crystal structure and chemical homogeneity of Tl2201 over the volume sampled by the transport measurements.

\subsection*{Resistivity measurements on cryo-FIB devices}

To establish that cryo-FIB devices yield reliable absolute transport measurements, we fabricated six-terminal $ab$-plane devices (see Methods) that enable simultaneous four-probe measurements of the in-plane resistivity and Hall effect (Fig.~\ref{fig3}a).

The absolute magnitude and temperature dependence of $\rho_{ab}(T)$ are in quantitative agreement with measurements on bulk single crystals (Fig.~\ref{fig3}b). These values are in good agreement with previous independent studies~\cite{Manako1992,Putzke2021} without any rescaling.

The Hall resistivity is linear in magnetic field above $T_c$ (Fig.~\ref{fig3}c), allowing reliable extraction of the Hall coefficient (see Methods). The corresponding Hall carrier density $n_{\rm H}$ agrees with previous measurements on bulk single crystals (Fig.~\ref{fig3}d)~\cite{Putzke2021}.

Together, these measurements demonstrate that cryo-FIB lamella-based microstructures can be fabricated with minimal damage and with no detectable degradation of the electronic properties. Once the small residual oxygen loss caused by local heating has been reversed by appropriate low-temperature annealing (see Methods), the devices display the same intrinsic properties as single crystals, with the additional advantages of well-defined geometries and precisely spaced electrical contacts.

Given the agreement of our $ab$-plane results with the bulk single-crystal literature, we expected the same level of agreement for $c$-axis transport. Instead, we found a systematic discrepancy with every previously reported measurement of $\rho_c$. We fabricated the devices using identical lamella extraction and annealing techniques to those for the $ab$-plane devices. The current path, channel dimensions, and contact separation were optimised for accurate determination of the absolute resistivity (Fig.~\ref{fig4}a and Methods). The superconducting $T_{\rm c}$ and the temperature dependence of the normal-state resistivity across a range of hole dopings (Fig.~\ref{fig4}b) are similar to previous findings on single crystals. The superconducting transition widths are typically $\Delta T_{\rm c}\sim1$--2~K, compared with $\Delta T_{\rm c}\gtrsim4$~K reported for comparable bulk single crystals~\cite{Ma2007,Manako1992,Hussey1996,Hussey2003,Herrmann1993}, consistent with the high quality and homogeneity of the microstructured devices. The absolute resistivity decreases monotonically with increasing hole doping, while the normal-state temperature dependence remains predominantly metallic across the full doping range studied, consistent with previous reports on Tl2201~\cite{Ma2007} and other overdoped cuprates~\cite{Honma2010}.

The discrepancy with the literature values obtained from bulk crystals concerns the absolute value of the resistivity. As shown in Fig.~\ref{fig4}c for highly overdoped samples with $T_{\rm c}\sim 13$\,K, the resistivity of the cryo-FIB device is systematically larger by a factor of approximately three over the entire temperature range. This is clearly not an issue of disorder: the residual resistivity ratio at $T_{\rm c}$ for the microstructured device ($\approx22$) is larger than that of the bulk crystal datasets (11.8--12.5).

To check this surprising observation carefully, we performed 26 measurements of $\rho_c(T)$ across nine devices and a range of $T_{\rm c}$ values (see Methods) and compared them against single-crystal data available in the literature~\cite{Ma2007,Hussey1996,Hussey2003,Manako1992,Herrmann1993}. The results are mutually consistent across all devices and dopings studied, demonstrating that the enhancement of $\rho_c$ is reproducible and is not associated with a particular device geometry or sample. At 200~K, the average resistivity enhancement relative to the equivalent single-crystal values is $2.9\pm0.3$.

\subsection*{Transport anisotropy and comparison with Boltzmann theory}

Our device transport data resolve a puzzling discrepancy between literature transport data and expectations based on the Fermi surface anisotropy deduced from measurements of quantum oscillations via the de Haas-van Alphen (dHvA) effect~\cite{Rourke2010,Bangura2010}. In Fig.~\ref{fig5}, we show the resistive anisotropy calculated using our cryo-FIB $c$-axis resistivity measurements and bulk in-plane resistivity data~\cite{Tyler1997,Tyler1998}. As shown in Fig.~\ref{fig3}b, our in-plane device data agree well in overall magnitude with the bulk data; however, because the bulk datasets include samples with $T_{\rm c}$ more closely matched (within $\sim$1--2 K) to those of our $c$-axis devices, we used the bulk data in calculating the anisotropy. Because our device $c$-axis resistivity is large, the measured anisotropy is considerably greater than that reported in Refs.~\cite{Manako1992,Hussey1996}.

Also shown in Fig.~\ref{fig5} are filled circles denoting the anisotropies deduced from Boltzmann calculations taking the Fermi surface warping established in dHvA measurements on crystals with similar $T_{\rm c}$, and assuming a single isotropic relaxation time. Our data provide an excellent match to the calculated anisotropy for the $T_{\rm c}\sim10$\,K sample and a plausible one to that for the $T_{\rm c}\sim30$\,K sample, for which an extrapolation with a larger temperature dependence would be required, as indicated by the dashed lines. We also note that the trend (decreasing anisotropy as $T_{\rm c}$ decreases) is well captured. In contrast, the temperature-dependent anisotropies reported in Ref.~\cite{Manako1992} and the low-temperature anisotropy reported in Ref.~\cite{Hussey1996} substantially underestimate the expectations based on the dHvA-derived Fermi surface warping.

%% file: Tex/discussion.tex
\section*{Discussion}

Combining cryo-FIB microstructuring with systematic transport measurements yields a quantitative and internally consistent picture of electronic transport in Tl2201. The agreement of our in-plane resistivity and Hall carrier density with bulk literature values indicates that cryo-FIB devices retain the intrinsic electronic properties of Tl2201, providing confidence that the $c$-axis measurements we report are quantitatively reliable.

The observed $\rho_c(T)$ is systematically larger than previously reported bulk values, an enhancement that is reproducible across devices of different dimensions and aspect ratios (Table~\ref{table2}). The discrepancy between the bulk-crystal and microstructure measurements was not anticipated. We speculate that it may arise from current short-circuiting through extended defects (for example, stacking faults or low-angle boundaries), whose influence would be expected to be reduced in microstructured devices because of the smaller sampled volume and better-defined current paths. Bulk measurements also average over larger volumes with poorly constrained geometric factors and contact areas, an uncertainty that microstructured devices largely avoid by construction. In addition, the four-terminal measurement region in the microstructured devices is defined entirely within a single-crystal lamella, with the metal contacts located outside the measured volume. The measured current distribution and voltage contact separation for the $c$-axis devices are therefore determined by the sculpted crystal geometry rather than by the crystal--metal interface, reducing uncertainties associated with current injection and contact geometry that are difficult to eliminate in conventional bulk measurements. While the present data do not allow these contributions to be individually isolated, the small scatter in the microstructure measurements in samples of similar $T_{\rm c}$ supports the hypothesis that microstructuring has greatly reduced, if not completely eliminated, the probability of encountering extended defects in any given device. Together, these factors provide a consistent explanation for why previous bulk measurements appear to have systematically underestimated the intrinsic $c$-axis resistivity of Tl2201.

The higher $\rho_c$ translates into a comparable increase in the resistive anisotropy $\rho_c/\rho_{ab}$, bringing it into agreement with the zero-temperature anisotropy calculated by Boltzmann transport theory using the experimentally established three-dimensional Fermi surface determined from quantum oscillation and AMRO experiments, together with an isotropic zero-temperature relaxation time approximation consistent with AMRO studies~\cite{Rourke2010,Bangura2010,Abdel-Jawad2006}. This agreement suggests that anisotropic scattering, or other hypotheses about the underlying physics, are not required to reconcile transport and Fermi-surface probes, thereby resolving a long-standing puzzle about out-of-plane transport in Tl2201.

Taken together with existing ARPES, quantum oscillation, AMRO, specific heat, and in-plane transport measurements~\cite{Wade1994,Carrington1996,Mackenzie1993,Mackenzie1996,Tyler1998,Hussey1996,Hussey2003,Plate2005,Abdel-Jawad2006,Analytis2007,Vignolle2008,Rourke2010,Bangura2010}, our results establish a quantitatively consistent picture of the normal-state electronic structure and transport of strongly overdoped Tl2201 that is unmatched in any other cuprate. The large Fermi surfaces satisfy Luttinger's theorem, and their experimentally established warping successfully describes the intrinsic $c$-axis transport. This, together with the long mean free paths, makes these strongly overdoped materials an ideal test-bed for theories studying the mechanism by which $T_{\rm c}$ is reduced by doping. Since the Fermi volume is large, the relative change in going from $T_{\rm c}=25$\,K to $T_{\rm c}=10$\,K is only a few per cent, so the large $T_{\rm c}$ change must be driven by a change in the strength of the pairing interaction. Successful prediction of this change should be a target for any microscopic theory of cuprate superconductivity.

Although our findings on Tl2201 microstructures are interesting in their own right, they have broader significance in demonstrating that the lamella cutting protocol can be applied to heat- and damage-sensitive materials. Our results open the way not only for reliable quantitative transport measurements, but more generally for the application of lamella-based microfabrication techniques to correlated quantum materials where FIB-induced damage has previously precluded such studies~\cite{Moll2018,Bachmann2019,Putzke2020,Diaz2022,Hunter2024}. We expect this work to pave the way toward a new generation of insight into some of the most fascinating materials and problems in condensed matter physics.

%% file: Tex/methods.tex
\section*{Methods}
\subsection*{Bulk sample preparation}
Single crystals of Tl2201 were grown using a self-flux method, and in-plane transport devices were fabricated following the procedures described in Ref.~\cite{Tyler1997,Tyler1998}.

\subsection*{Cryo-FIB microstructuring}
Table~\ref{table1} summarises the processing parameters used to fabricate Tl2201 transport devices using a cryo-Xe$^+$ FIB. Devices were also fabricated using a Ga$^+$ FIB with a Kleindiek MHCS stage attachment, cooling samples to $-60^\circ$C.

\begin{enumerate}
    \item {\bf Initial preparation and orientation.} Single crystals of Tl2201 were oxygen-annealed to increase the interstitial oxygen content. Samples were mounted on SEM stubs and oriented using EBSD prior to lamella extraction along target crystallographic directions.
    
    \item {\bf Cryogenic lamella fabrication.} Samples were prepared using a ThermoFisher Helios 4 PFIB UXe equipped with an Aquilos cryostage. All milling and polishing steps were performed at $-190\,^\circ$C. Rectangular lamellae were defined along the desired crystallographic directions with an initial thickness of ${\sim}10\,$\textmu{}m and typical lateral dimensions of $30 \times 100~$\textmu{}m for $c$-axis devices or $200 \times 50~$\textmu{}m for $ab$-plane devices, then transferred \textit{in situ} onto a TEM grid.
    
    \item {\bf Lamella thinning and shaping.} Lamellae were thinned to a final thickness of ${\sim}5\,$\textmu{}m, with stage tilts adjusted to keep surfaces parallel to the desired crystallographic planes. Sloped contact edges were introduced at $45^\circ$ through the full lamella thickness, followed by a final low-energy (2~kV) polishing step from multiple orientations.
    
    \item {\bf Surface preparation and metallisation.} Samples were transferred ex situ to a vacuum chamber equipped with a Kaufman ion source. Surface contaminants were removed by Ar ion etching, followed by in situ Au deposition. The combination of Ar ion cleaning and high-power Au deposition consistently produced reproducible ohmic contacts in the 1--10\,$\Omega$ range on all devices studied.
    
    \item {\bf Device assembly and contacting.} Lamellae were returned to the FIB-SEM, where a protective Pt layer was deposited onto the metallised contact edges. Samples were transferred and integrated onto Au-coated sapphire substrates, with mechanical support and electrical contacts established via FIB-induced Pt deposition.
    
    \item {\bf Final patterning.} Both the lamella and substrate were patterned at cryogenic temperature to define the device geometries shown in Fig.~\ref{fig3}a and Fig.~\ref{fig4}a. The voltage contact separation $l$, channel width $w$, and thickness $t$ were extracted from SEM images of the fabricated devices. Typical device dimensions are given in Table~\ref{table2}.

\end{enumerate}

\subsection*{Structural characterisation}
Single-crystal XRD on as-grown crystals and cryo-FIB lamellae was performed using a Rigaku XtaLAB Synergy-DW diffractometer with CrysAlisPro 171.42.79a software, used to index diffraction datasets and generate precession images.

Lamellae for scanning transmission electron microscopy (STEM) were prepared using the standard FIB lift-out procedure on the same Thermo Fisher Helios 3 Ga$^+$ FIB with Kleindiek MHCS stage attachment. All trenching, polishing, and thinning steps were performed with the sample cooled to $-60\,^\circ$C; deposition of surface protective layers and lift-out by micromanipulator were performed at room temperature. Electron microscopy measurements were performed under cryogenic cooling using a DENSSolutions Lightning Arctic dual-tilt sample holder cooled with liquid nitrogen to reduce the possibility of beam damage. Atomic-resolution STEM measurements were performed using a probe-corrected Thermo Spectra X-FEG microscope operated at 300~kV with a 75~\textmu{}m C2 aperture, with images acquired along the [100] zone axis.

\subsection*{Oxygen annealing}

Oxygen annealing was performed in a Heraeus ROF7/50 tube furnace with a Eurotherm 2404 controller. A quartz tube connected to gas inlets and vacuum lines enabled stable oxygen pressures between $0.001$ and $1\,\mathrm{bar}$ for up to $30\,\mathrm{h}$. Single crystals, mounted in a Pt boat, were placed in the quartz tube at $450\,^\circ\mathrm{C}$, evacuated to $10^{-6}\,\mathrm{bar}$, and purged with $1\,\mathrm{bar}$ of oxygen $3$--$4$ times, before being filled with a mixture of oxygen ($0.001$--$1\,\mathrm{bar}$) and $1\,\mathrm{bar}$ Ar. Samples were equilibrated for $8$--$24\,\mathrm{h}$ and quenched to room temperature within approximately $45\,\mathrm{s}$ by removing the Pt boat from the hot zone and transferring it to a room-temperature metal block under flowing Ar.

For vacuum anneals ($T_\mathrm{c} > 55\,\mathrm{K}$), the same furnace and Pt boat were used, with the quartz tube continuously evacuated by a turbomolecular pump to $\sim 10^{-8}\,\mathrm{bar}$ while samples were held between $300$ and $500\,^\circ\mathrm{C}$ for $4$--$12\,\mathrm{h}$, followed by the same quench procedure.

Sapphire-mounted microstructured devices deform on heating above $300\,^\circ$C and fail mechanically below $450\,^\circ$C. Low-temperature protocols at $250\,^\circ$C, using the same low-pressure oxygen and vacuum annealing conditions, were developed for the devices.

A non-equilibrium annealing strategy was adopted for the devices, quenching before equilibrium was reached: repeated annealing of a single device under identical conditions showed $T_\mathrm{c}$ stabilising after 25--30~h of cumulative annealing at $250\,^\circ$C, consistent with the bulk single-crystal annealing timescales reported by Tyler \textit{et al.}~\cite{Tyler1997}; full equilibration would access only $T_{\rm c} = 13$--36~K. Each anneal was performed in multiple steps at static oxygen pressure, optimised for the initial composition and desired final doping. The non-equilibrium annealing parameters used in this work are listed in Table~\ref{table3}.

\subsection*{Electrical transport measurements}
The bulk crystal data shown in Fig.~\ref{fig3}b were obtained using methods as described in Ref.~\cite{Tyler1997}. Microstructure transport measurements were performed on the two device geometries shown in Fig.~\ref{fig3}--\ref{fig4}: four-point longitudinal devices for $c$-axis transport, and six-point Hall bar devices for $ab$-plane transport and Hall effect. Samples were mounted on Quantum Design transport pucks and inserted into a PPMS for measurements between 2--300~K at 0--14~T.

An AC current $I_x$ in the range 1--10~\textmu{}A was applied between the current contacts, and the resulting voltage drops $V_x$ and $V_{\rm H}$ were recorded using a SynkTek MCL1-540 multi-channel lock-in amplifier system. Simultaneous acquisition of demodulated lock-in outputs and full time-domain waveforms allowed reconstruction of cycle-resolved current--voltage characteristics at sufficiently low excitation frequencies. The measurement frequency was chosen in the range 2--100~Hz to avoid mains pickup and $1/f$ noise while maintaining an undistorted $I$--$V$ response.

The longitudinal resistivity along the $a$- or $c$-axis is given by
\begin{equation}
    \rho_{x} = \frac{V_{x}}{I_x}\frac{w t}{l},
\end{equation}
where $x = a$ or $c$.

For Hall bar devices, the Hall resistivity $\rho_{_{{\mathrm{H}}}}$ and Hall number $n_{_{{\mathrm{H}}}}$ are defined as
\begin{equation}
    \rho_{_{{\mathrm{H}}}} = \frac{V_{_{{\mathrm{H}}}} t}{I_a},
\end{equation}
\begin{equation}
    n_{_{{\mathrm{H}}}} = \frac{B}{e \rho_{_{{\mathrm{H}}}}}.
\end{equation}

The Hall voltage was antisymmetrised with respect to magnetic field using sweeps between $-10$ and $+10$~T to remove longitudinal contributions from contact misalignment (typically 1--2\% in the microstructure devices).

\subsection*{Magnetic susceptibility measurements}
Samples were mounted on quartz holders using Apiezon N grease and characterised by AC magnetic susceptibility measurements in a 14\,T PPMS or a 7\,T MPMS3 system from Quantum Design. The superconducting transition was measured using a $177\,\mathrm{Hz}$ AC drive with an amplitude of $1\,\mathrm{Oe}$ applied along $H \parallel c$, with a small DC magnetic field offset ($\sim 0.2$~mT) applied to minimise the effects of residual and stray magnetic fields. The transition width $\Delta T_\mathrm{c}$ was defined as the temperature interval between 20\% and 80\% of the jump in the in-phase susceptibility signal; the overall transition shape is additionally influenced by disorder, demagnetisation effects, and variations in the superfluid density.

%% file: Tex/supplement.tex
\begin{figure*}[b]
\centering
\includegraphics[scale=0.82]{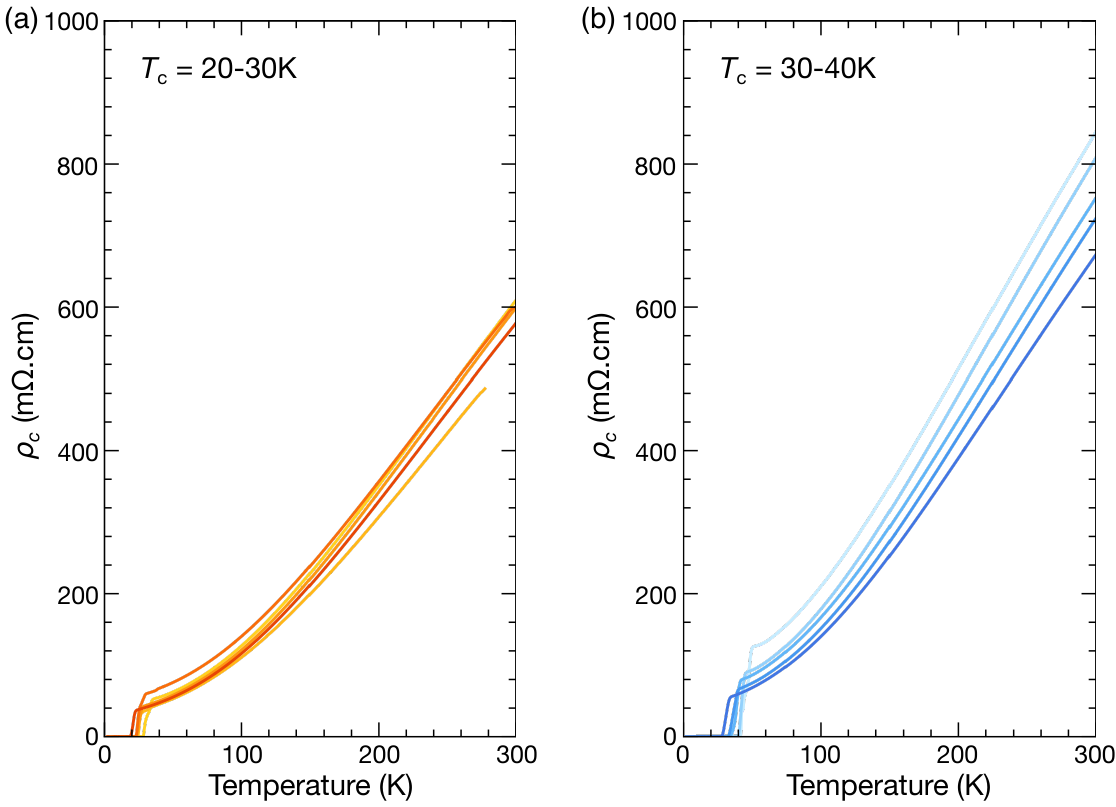}
\caption{{\bf Comparison of $c$-axis microstructured devices} with $T_{\rm c}$ values in the ranges (a) 20--30\,K and (b) 30--40\,K. The variation between the datasets reflects both the systematic doping dependence of $\rho_c(T)$ associated with differences in $T_{\rm c}$ and the experimental uncertainties of the measurements. For devices with comparable $T_{\rm c}$, the absolute $\rho_c(T)$ values are in good agreement, demonstrating the reproducibility of the microstructured device measurements.}
\label{figA1}
\end{figure*}

\begin{table*}[t]
\caption{Key FIB processing parameters used for Tl2201 device fabrication. Milling is performed under cryogenic conditions to suppress beam-induced heating and oxygen mobility; deposition steps are carried out at room temperature.} \label{table1}
\begin{ruledtabular}
\begin{tabular}{lcll}
{\bf Process step} & {\bf Temperature} & {\bf Parameters} & {\bf Tool} \\
\colrule
Lamella definition         & $-190\,^\circ$C & 30~kV, 60~nA & Xe$^+$ FIB\\
Thinning / polishing       & $-190\,^\circ$C & 30~kV, 4~nA & Xe$^+$ FIB\\
Surface cleaning           & $-190\,^\circ$C & 2~kV, 10~nA & Xe$^+$ FIB\\
Final device patterning    & $-190\,^\circ$C & 30~kV, 1~nA & Xe$^+$ FIB\\
\colrule
Pt deposition (transfer)   & $25\,^\circ$C   & 30~kV, 4~nA & Xe$^+$ FIB\\
Pt deposition (protection) & $25\,^\circ$C   & 16~kV, 0.43~nA & Xe$^+$ FIB\\
Pt deposition (contacts)   & $25\,^\circ$C   & 30~kV, 1~nA & Xe$^+$ FIB\\
\colrule
Ar etching                 & $25\,^\circ$C   & 0.2~nm/min, 30~min & Kaufman source \\
Au deposition              & $25\,^\circ$C   & 1~kW, 30~s sputtering & Kaufman source \\
\end{tabular}
\end{ruledtabular}
\end{table*}

\begin{table*}[t]
\begin{ruledtabular}
\caption{Physical dimensions and transport measurements for the Tl2201 microstructured devices investigated in this work. Multiple entries for a given device correspond to measurements performed after successive annealing steps at different superconducting transition temperatures $T_{\rm c}$. The range of device sizes, aspect ratios, and measured dopings demonstrates that the reported transport properties are reproducible across different microstructures and are not dominated by geometric artefacts.}
\label{table2}
\centering
\setlength{\tabcolsep}{16pt}
\begin{tabular}{c c | c c c | c c}
    \multicolumn{2}{c|}{\bf Samples} &\multicolumn{3}{c|}{\bf Dimensions} &\multicolumn{2}{c}{\bf Measurements}\\
    {Device} & {Category} & {Length} & {Width} & {Thickness} & {$T_c$} & {$\rho$(200\,K)}\\
    \colrule
    STRC00265 & $ab$-plane & 32.4\,\textmu{}m & 3.1\,\textmu{}m & 1.5\,\textmu{}m
        & 80\,K & 252\,\textmu{}$\Omega$.cm\\
    AM-abdev1 & $ab$-plane  & 10.5\,\textmu{}m & 4.5\,\textmu{}m & 2.5\,\textmu{}m
        & 60\,K & 143\,\textmu{}$\Omega$.cm\\
        
    AM-abdev2 & $ab$-plane  & 8.0\,\textmu{}m & 4.0\,\textmu{}m & 2.0\,\textmu{}m
        & 54\,K & 141\,\textmu{}$\Omega$.cm\\
    
    AM-abdev5 & $ab$-plane  & 19.9\,\textmu{}m & 2.3\,\textmu{}m & 2.7\,\textmu{}m
        & 66\,K & 168\,\textmu{}$\Omega$.cm\\
    &&&&& 89\,K & 205\,\textmu{}$\Omega$.cm\\
    
    \colrule
    
    CRYS00406 & $c$-axis  & 7.5\,\textmu{}m & 2.2\,\textmu{}m & 3.4\,\textmu{}m 
        & 25\,K & 357\,m$\Omega$.cm\\
    &&&&& 45\,K & 514\,m$\Omega$.cm\\
    &&&&& 71\,K & 700\,m$\Omega$.cm\\
    &&&&& 81\,K & 823\,m$\Omega$.cm\\
    
    AM-acdev1 & $c$-axis & 7.8\,\textmu{}m & 10.0\,\textmu{}m & 6.9\,\textmu{}m
        & 82\,K & 1014\,m$\Omega$.cm\\
    
    AM-acdev3 & $c$-axis & 2.5\,\textmu{}m & 1.8\,\textmu{}m & 1.2\,\textmu{}m
        & 79\,K & 815\,m$\Omega$.cm\\
    
    AM-acdev4 & $c$-axis & 6.7\,\textmu{}m & 4.8\,\textmu{}m & 4.9\,\textmu{}m
        & 13\,K & 292\,m$\Omega$.cm\\
    &&&&& 21\,K & 329\,m$\Omega$.cm\\
    &&&&& 25\,K & 342\,m$\Omega$.cm\\
    &&&&& 31\,K & 390\,m$\Omega$.cm\\
    &&&&& 36\,K & 419\,m$\Omega$.cm\\
    &&&&& 39\,K & 443\,m$\Omega$.cm\\
    &&&&& 42\,K & 478\,m$\Omega$.cm\\
    &&&&& 52\,K & 529\,m$\Omega$.cm\\
    &&&&& 62\,K & 602\,m$\Omega$.cm\\
    &&&&& 72\,K & 664\,m$\Omega$.cm\\
    &&&&& 80\,K & 750\,m$\Omega$.cm\\
    
    AM-acdev5 & $c$-axis & 10.8\,\textmu{}m & 2.7\,\textmu{}m & 4.9\,\textmu{}m
        & 68\,K & 621\,m$\Omega$.cm\\
    &&&&& 80\,K & 737\,m$\Omega$.cm\\
    
    AMLH-dev7 & $c$-axis & 8.7\,\textmu{}m & 4.0\,\textmu{}m & 3.5\,\textmu{}m
        & 29\,K & 340\,m$\Omega$.cm\\
    &&&&& 31\,K & 351\,m$\Omega$.cm\\
    
    AMLH-dev8 & $c$-axis & 9.5\,\textmu{}m & 5.0\,\textmu{}m & 3.4\,\textmu{}m
        & 17\,K & 289\,m$\Omega$.cm\\
    &&&&& 25\,K & 307\,m$\Omega$.cm\\
    &&&&& 86\,K & 1592\,m$\Omega$.cm\\
    
    AMLH-dev9 & $c$-axis & 8.6\,\textmu{}m & 5.7\,\textmu{}m & 2.9\,\textmu{}m
        & 67\,K & 643\,m$\Omega$.cm\\
    
    AMLH-dev10 & $c$-axis & 9.6\,\textmu{}m & 4.2\,\textmu{}m & 3.9\,\textmu{}m
        & 35\,K & 375\,m$\Omega$.cm\\
\end{tabular}
\end{ruledtabular}
\end{table*}

\begin{table*}[t]
\caption{Annealing protocol used to tune the hole doping of Tl2201 microstructured devices. Each annealing step starts from the state produced by the previous treatment. The initial oxygen anneal establishes a reference composition by proceeding to equilibrium, while subsequent anneals progressively reduce the oxygen content and increase $T_{\rm c}$. The table also lists the number of devices measured at each doping and the corresponding average $c$-axis resistivity at 200\,K.} \label{table3}
\begin{ruledtabular}
\begin{tabular}{c | c c c c | c c}
{\bf Hole doping~~}&\multicolumn{4}{c|}{\bf Annealing recipe}&\multicolumn{2}{c}{\bf Transport measurements}\\
{$T_{\rm c}$} & {Initial $T_{\rm c}$} & {Oxygen pressure} & {Temperature} & {Duration~~} & {Samples studied} & $\rho_c$(200\,K)\\
 \colrule
 15\,K & As fabricated & 1~bar & 250~$^\circ$C & 30~h& 2 & $290\pm2$\,m$\Omega$.cm\\
 20\,K & 10\,K & 0.002~bar  & 250~$^\circ$C & 16~h & 2 & $318\pm11$\,m$\Omega$.cm\\
 30\,K & 20\,K & 0.002~bar  & 250~$^\circ$C & 16~h& 3 & $350\pm4$\,m$\Omega$.cm\\
 35\,K & 30\,K & 0.002~bar  & 250~$^\circ$C & 8~h& 3 & $418\pm16$\,m$\Omega$.cm\\
 40\,K & 35\,K & $<10^{-8}$~bar & 250~$^\circ$C & 6~h& 2 & $496\pm18$\,m$\Omega$.cm\\
 50\,K & 40\,K & $<10^{-8}$~bar & 250~$^\circ$C & 8~h& 1 & $529$\,m$\Omega$.cm\\
 60\,K & 50\,K & $<10^{-8}$~bar & 250~$^\circ$C & 5~h& 1 & $602$\,m$\Omega$.cm\\
 70\,K & 60\,K & $<10^{-8}$~bar & 250~$^\circ$C & 5~h& 4 & $657\pm17$\,m$\Omega$.cm\\
 80\,K & 70\,K & $<10^{-8}$~bar & 250~$^\circ$C & 5~h& 4 & $781\pm20$\,m$\Omega$.cm\\
 82\,K & 80\,K & $<10^{-8}$~bar & 250~$^\circ$C & 5~h& 1 & $1014$\,m$\Omega$.cm\\
 86\,K & 82\,K & $<10^{-8}$~bar & 300~$^\circ$C & 5~h& 1 & $1592$\,m$\Omega$.cm\\
\end{tabular}
\end{ruledtabular}
\end{table*}

%% file: Tex/references.bib
@article{Keimer2015,
	author = {Keimer, B. and Kivelson, S. A. and Norman, M. R. and Uchida, S. and Zaanen, J.},
	doi = {10.1038/nature14165},
	id = {Keimer2015},
	issn = {1476-4687},
	journal = {Nature},
	number = {7538},
	pages = {179-186},
	title = {From quantum matter to high-temperature superconductivity in copper oxides},
	url = {https://doi.org/10.1038/nature14165},
	volume = {518},
	year = {2015}}

@incollection{Ekin2006,
    author = {Ekin, J. W.},
    isbn = {9780198570547},
    title = {Sample contacts},
    booktitle = {Experimental Techniques for Low-Temperature Measurements: Cryostat Design, Material Properties and Superconductor Critical-Current Testing},
    publisher = {Oxford University Press},
    year = {2006},
    month = {10},
    doi = {10.1093/acprof:oso/9780198570547.003.0008}}

@ARTICLE{Alloul2024,
AUTHOR={Alloul, H.},   
TITLE={What do we learn from impurities and disorder in high-$T_{\rm c}$ cuprates?},
JOURNAL={Frontiers in Physics},
VOLUME={12},
pages={1406242},
ISSN={2296-424X},
YEAR={2024},
URL={https://www.frontiersin.org/journals/physics/articles/10.3389/fphy.2024.1406242},
DOI={10.3389/fphy.2024.1406242}}

@article{Moll2018,
   author = "Moll, P. J.W.",
   title = "Focused Ion Beam microstructuring of quantum matter", 
   journal= "Annual Review of Condensed Matter Physics",
   year = "2018",
   volume = "9",
   number = "Volume 9, 2018",
   pages = "147-162",
   doi = "10.1146/annurev-conmatphys-033117-054021",
   url = "https://www.annualreviews.org/content/journals/10.1146/annurev-conmatphys-033117-054021",
   publisher = "Annual Reviews",
   issn = "1947-5462",
   type = "Journal Article"}

@phdthesis{Bachmann2019,
  author = {Bachmann, Maja Deborah},
  title = {Manipulating Anisotropic Transport and Superconductivity by Focused Ion Beam Microstructuring},
  school = {University of St Andrews},
  year = {2019},
  doi = {10.17630/10023-17866},
  url = {https://hdl.handle.net/10023/17866}}

@article{Kushwaha2017,
author = {Kushwaha, P. and Borrmann, H. and Khim, S. and Rosner, H. and Moll, P. J. W. and Sokolov, D. A. and Sunko, V. and Grin, Yu. and Mackenzie, A. P.},
title = {Single Crystal Growth, Structure, and Electronic Properties of Metallic Delafossite PdRhO2},
journal = {Crystal Growth \& Design},
volume = {17},
number = {8},
pages = {4144-4150},
year = {2017},
doi = {10.1021/acs.cgd.7b00418},
URL = {https://doi.org/10.1021/acs.cgd.7b00418}}

@article{Putzke2020,
author = {C. Putzke  and M. D. Bachmann  and P. McGuinness  and E. Zhakina  and V. Sunko  and M. Konczykowski  and T. Oka  and R. Moessner  and A. Stern  and M. König  and S. Khim  and A. P. Mackenzie  and P. J. W. Moll },
title = {$h/e$ oscillations in interlayer transport of delafossites},
journal = {Science},
volume = {368},
number = {6496},
pages = {1234-1238},
year = {2020},
doi = {10.1126/science.aay8413},
URL = {https://www.science.org/doi/abs/10.1126/science.aay8413}}

@article{Wagner1997,
title = {Multiple defects in overdoped Tl$_2$Ba$_2$CuO$_{6+\delta}$: Effects on structure and superconductivity},
journal = {Physica C: Superconductivity},
volume = {277},
number = {3},
pages = {170-182},
year = {1997},
issn = {0921-4534},
doi = {10.1016/S0921-4534(97)00062-2},
url = {https://www.sciencedirect.com/science/article/pii/S0921453497000622},
author = {J. L. Wagner and O. Chmaissem and J. D. Jorgensen and D. G. Hinks and P. G. Radaelli and B. A. Hunter and W. R. Jensen}}

@phdthesis{Peets2008,
  author       = {Darren C. Peets},
  title        = {Why be normal? Single crystal growth and the startlingly unremarkable electronic structure of Tl2201},
  school       = {The University of British Columbia, Vancouver},
  year         = {2008},
  month        = {07},
  day          = {17},
  doi          = {10.14288/1.0066236},
  url          = {https://www.jick.net/theses/Peets/thesis.pdf}}

@article{Shimakawa1993,
title = {Chemical and structural study of tetragonal and orthorhombic Tl$_2$Ba$_2$CuO$_6$},
journal = {Physica C: Superconductivity},
volume = {204},
number = {3},
pages = {247-261},
year = {1993},
issn = {0921-4534},
doi = {10.1016/0921-4534(93)91006-H},
url = {https://www.sciencedirect.com/science/article/pii/092145349391006H},
author = {Y. Shimakawa}}

@article{Sefrioui2001,
  title = {Vortex liquid entanglement in irradiated ${\mathrm{YBa}}_{2}{\mathrm{Cu}}_{3}{\mathrm{O}}_{7}$ thin films},
  author = {Sefrioui, Z. and Arias, D. and Gonz\'alez, E. M. and Le\'on, C. and Santamaria, J. and Vicent, J. L.},
  journal = {Physical Review B},
  volume = {63},
  issue = {6},
  pages = {064503},
  numpages = {5},
  year = {2001},
  month = {01},
  publisher = {American Physical Society},
  doi = {10.1103/PhysRevB.63.064503},
  url = {https://link.aps.org/doi/10.1103/PhysRevB.63.064503}}

@Article{Caruso2023,
AUTHOR = {Caruso, R. and Camino, F. and Gu, G. and Tranquada, J. M. and Han, M. G. and Zhu, Y. and Bollinger, A. T. and Božović, I.},
TITLE = {Effects of Focused Ion Beam lithography on La$_{2-x}$Sr$_x$CuO$_4$ single crystals},
JOURNAL = {Condensed Matter},
VOLUME = {8(2)},
YEAR = {2023},
pages = {35},
URL = {https://www.mdpi.com/2410-3896/8/2/35},
ISSN = {2410-3896},
DOI = {10.3390/condmat8020035}}

@article{Hensel1997,
title = {Defects in high-$T_{\rm c}$ superconductors after ion irradiation},
journal = {Journal of Nuclear Materials},
volume = {251},
pages = {218-224},
year = {1997},
issn = {0022-3115},
doi = {10.1016/S0022-3115(97)00225-0},
url = {https://www.sciencedirect.com/science/article/pii/S0022311597002250},
author = {B. Hensel}}

@article{Wade1994,
	author = {Wade, J. M. and Loram, J. W. and Mirza, K. A. and Cooper, J. R. and Tallon, J. L.},
	doi = {10.1007/BF00730408},
	id = {Wade1994},
	issn = {1572-9605},
	journal = {Journal of Superconductivity},
	number = {1},
	pages = {261-264},
	title = {Electronic specific heat of Tl$_2$Ba$_2$CuO$_{6+\delta}$ from 2~K to 300~K for $0\le\delta\le0.1$},
	url = {https://doi.org/10.1007/BF00730408},
	volume = {7},
	year = {1994}}

@article{Carrington1996,
  title = {Specific heat of low-${T}_\mathrm{c}$ ${\mathrm{Tl}}_{2}{\mathrm{Ba}}_{2}\mathrm{Cu}{\mathrm{O}}_{6+\ensuremath{\delta}}$},
  author = {Carrington, A. and Mackenzie, A. P. and Tyler, A.},
  journal = {Physical Review B},
  volume = {54},
  issue = {6},
  pages = {R3788-R3791},
  numpages = {0},
  year = {1996},
  month = {08},
  publisher = {American Physical Society},
  doi = {10.1103/PhysRevB.54.R3788},
  url = {https://link.aps.org/doi/10.1103/PhysRevB.54.R3788}}

@article{Mackenzie1993,
  title = {Resistive upper critical field of ${\mathrm{Tl}}_{2}$${\mathrm{Ba}}_{2}$${\mathrm{CuO}}_{6}$ at low temperatures and high magnetic fields},
  author = {Mackenzie, A. P. and Julian, S. R. and Lonzarich, G. G. and Carrington, A. and Hughes, S. D. and Liu, R. S. and Sinclair, D. S.},
  journal = {Physical Review Letters},
  volume = {71},
  issue = {8},
  pages = {1238-1241},
  numpages = {0},
  year = {1993},
  month = {08},
  publisher = {American Physical Society},
  doi = {10.1103/PhysRevLett.71.1238},
  url = {https://link.aps.org/doi/10.1103/PhysRevLett.71.1238}}

@article{Mackenzie1996,
  title = {Normal-state magnetotransport in superconducting ${\mathrm{Tl}}_{2}$${\mathrm{Ba}}_{2}$${\mathrm{CuO}}_{6+\mathrm{\ensuremath{\delta}}}$ to millikelvin temperatures},
  author = {Mackenzie, A. P. and Julian, S. R. and Sinclair, D. C. and Lin, C. T.},
  journal = {Physical Review B},
  volume = {53},
  issue = {9},
  pages = {5848--5855},
  numpages = {0},
  year = {1996},
  month = {Mar},
  publisher = {American Physical Society},
  doi = {10.1103/PhysRevB.53.5848},
  url = {https://link.aps.org/doi/10.1103/PhysRevB.53.5848}}

@article{Tyler1998,
  title = {High-field study of normal-state magnetotransport in ${\mathrm{Tl}}_{2}{\mathrm{Ba}}_{2}{\mathrm{CuO}}_{6+\mathrm{\ensuremath{\delta}}}$},
  author = {Tyler, A. W. and Ando, Y. and Balakirev, F. F. and Passner, A. and Boebinger, G. S. and Schofield, A. J. and Mackenzie, A. P. and Laborde, O.},
  journal = {Physical Review B},
  volume = {57},
  issue = {2},
  pages = {R728-R731},
  numpages = {0},
  year = {1998},
  month = {01},
  publisher = {American Physical Society},
  doi = {10.1103/PhysRevB.57.R728},
  url = {https://link.aps.org/doi/10.1103/PhysRevB.57.R728}}

@article{Hussey1996,
  title = {Angular dependence of the $c$-axis normal state magnetoresistance in single crystal ${\mathrm{Tl}}_{2}{\mathrm{Ba}}_{2}{\mathrm{CuO}}_{6}$},
  author = {Hussey, N. E. and Cooper, J. R. and Wheatley, J. M. and Fisher, I. R. and Carrington, A. and Mackenzie, A. P. and Lin, C. T. and Milat, O.},
  journal = {Physical Review Letters},
  volume = {76},
  issue = {1},
  pages = {122-125},
  numpages = {0},
  year = {1996},
  month = {01},
  publisher = {American Physical Society},
  doi = {10.1103/PhysRevLett.76.122},
  url = {https://link.aps.org/doi/10.1103/PhysRevLett.76.122}}

@Article{Hussey2003,
author={Hussey, N. E.
and Abdel-Jawad, M.
and Carrington, A.
and Mackenzie, A. P.
and Balicas, L.},
title={A coherent three-dimensional Fermi surface in a high-transition-temperature superconductor},
journal={Nature},
year={2003},
month={10},
day={01},
volume={425},
number={6960},
pages={814-817},
issn={1476-4687},
doi={10.1038/nature01981},
url={https://doi.org/10.1038/nature01981}}

@article{Plate2005,
  title = {Fermi surface and quasiparticle excitations of overdoped ${\mathrm{Tl}}_{2}{\mathrm{Ba}}_{2}{\mathrm{CuO}}_{6+\ensuremath{\delta}}$},
  author = {Plat\'e, M. and Mottershead, J. D. F. and Elfimov, I. S. and Peets, D. C. and Liang, Ruixing and Bonn, D. A. and Hardy, W. N. and Chiuzbaian, S. and Falub, M. and Shi, M. and Patthey, L. and Damascelli, A.},
  journal = {Physical Review Letters},
  volume = {95},
  issue = {7},
  pages = {077001},
  numpages = {4},
  year = {2005},
  month = {08},
  publisher = {American Physical Society},
  doi = {10.1103/PhysRevLett.95.077001},
  url = {https://link.aps.org/doi/10.1103/PhysRevLett.95.077001}}

@Article{Abdel-Jawad2006,
author={Abdel-Jawad, M.
and Kennett, M. P.
and Balicas, L.
and Carrington, A.
and Mackenzie, A. P.
and McKenzie, R. H.
and Hussey, N. E.},
title={Anisotropic scattering and anomalous normal-state transport in a high-temperature superconductor},
journal={Nature Physics},
year={2006},
month={12},
day={01},
volume={2},
number={12},
pages={821-825},
issn={1745-2481},
doi={10.1038/nphys449},
url={https://doi.org/10.1038/nphys449}}

@Article{Vignolle2008,
author={Vignolle, B.
and Carrington, A.
and Cooper, R. A.
and French, M. M. J.
and Mackenzie, A. P.
and Jaudet, C.
and Vignolles, D.
and Proust, C.
and Hussey, N. E.},
title={Quantum oscillations in an overdoped high-$T_\mathrm{c}$ superconductor},
journal={Nature},
year={2008},
month={10},
day={01},
volume={455},
number={7215},
pages={952-955},
issn={1476-4687},
doi={10.1038/nature07323},
url={https://doi.org/10.1038/nature07323}}

@article{Rourke2010,
doi = {10.1088/1367-2630/12/10/105009},
url = {https://doi.org/10.1088/1367-2630/12/10/105009},
year = {2010},
month = {10},
publisher = {},
volume = {12},
number = {10},
pages = {105009},
author = {Rourke, P. M. C. and Bangura, A. F. and Benseman, T. M. and Matusiak, M. and Cooper, J. R. and Carrington, A. and Hussey, N. E.},
title = {A detailed de Haas-van Alphen effect study of the overdoped cuprate Tl$_2$Ba$_2$CuO$_{6+\delta}$},
journal = {New Journal of Physics}}

@article{Bangura2010,
  title = {Fermi surface and electronic homogeneity of the overdoped cuprate superconductor Tl$_{2}$Ba$_{2}$CuO$_{6+\ensuremath{\delta}}$ as revealed by quantum oscillations},
  author = {Bangura, A. F. and Rourke, P. M. C. and Benseman, T. M. and Matusiak, M. and Cooper, J. R. and Hussey, N. E. and Carrington, A.},
  journal = {Physical Review B},
  volume = {82},
  issue = {14},
  pages = {140501},
  numpages = {4},
  year = {2010},
  month = {10},
  publisher = {American Physical Society},
  doi = {10.1103/PhysRevB.82.140501},
  url = {https://link.aps.org/doi/10.1103/PhysRevB.82.140501}}

@phdthesis{Tyler1997,
  author       = {Tyler, A. W.},
  title        = {An investigation into the magnetotransport properties of layered superconducting perovskites},
  school       = {University of Cambridge},
  year         = {1997}}

@article{Manako1992,
  title = {Transport and structural study of ${\mathrm{Tl}}_{2}$${\mathrm{Ba}}_{2}$${\mathrm{CuO}}_{6+\mathrm{\ensuremath{\delta}}}$ single crystals prepared by the KCl flux method},
  author = {Manako, T. and Kubo, Y. and Shimakawa, Y.},
  journal = {Physical Review B},
  volume = {46},
  issue = {17},
  pages = {11019-11024},
  numpages = {0},
  year = {1992},
  month = {11},
  publisher = {American Physical Society},
  doi = {10.1103/PhysRevB.46.11019},
  url = {https://link.aps.org/doi/10.1103/PhysRevB.46.11019}}

@Article{Putzke2021,
author={Putzke, C.
and Benhabib, S.
and Tabis, W.
and Ayres, J.
and Wang, Z.
and Malone, L.
and Licciardello, S.
and Lu, J.
and Kondo, T.
and Takeuchi, T.
and Hussey, N. E.
and Cooper, J. R.
and Carrington, A.},
title={Reduced Hall carrier density in the overdoped strange metal regime of cuprate superconductors},
journal={Nature Physics},
year={2021},
month={07},
day={01},
volume={17},
number={7},
pages={826-831},
issn={1745-2481},
doi={10.1038/s41567-021-01197-0},
url={https://doi.org/10.1038/s41567-021-01197-0}}

@article{Rullier-Albenque2001,
  title = {Disorder and transport in cuprates: Weak localization and magnetic contributions},
  author = {Rullier-Albenque, F. and Alloul, H. and Tourbot, R.},
  journal = {Physical Review Letters},
  volume = {87},
  issue = {15},
  pages = {157001},
  numpages = {4},
  year = {2001},
  month = {09},
  publisher = {American Physical Society},
  doi = {10.1103/PhysRevLett.87.157001},
  url = {https://link.aps.org/doi/10.1103/PhysRevLett.87.157001}}

@article{Kim2011,
title = {Minimization of focused ion beam damage in nanostructured polymer thin films},
journal = {Ultramicroscopy},
volume = {111},
number = {3},
pages = {191-199},
year = {2011},
issn = {0304-3991},
doi = {10.1016/j.ultramic.2010.11.027},
url = {https://www.sciencedirect.com/science/article/pii/S0304399110003177},
author = {S. Kim and M. J. Park and N. P. Balsara and G. Liu and A. M. Minor}}

@article{Narayan2015,
	author = {Narayan, K. and Subramaniam, S.},
	doi = {10.1038/nmeth.3623},
	id = {Narayan2015},
	issn = {1548-7105},
	journal = {Nature Methods},
	number = {11},
	pages = {1021-1031},
	title = {Focused Ion Beams in biology},
	url = {https://doi.org/10.1038/nmeth.3623},
	volume = {12},
	year = {2015}}

@article{Noble2024,
title = {Cryo-Focused Ion Beam for \textit{in situ} structural biology: State of the art, challenges, and perspectives},
journal = {Current Opinion in Structural Biology},
volume = {87},
pages = {102864},
year = {2024},
issn = {0959-440X},
doi = {10.1016/j.sbi.2024.102864},
url = {https://www.sciencedirect.com/science/article/pii/S0959440X24000915},
author = {A. J. Noble and A. de Marco}}

@article{Ma2007,
doi = {10.1088/0953-8984/19/18/186203},
url = {https://doi.org/10.1088/0953-8984/19/18/186203},
year = {2007},
month = {04},
publisher = {},
volume = {19},
number = {18},
pages = {186203},
author = {Ma, Y. C. and Liu, J. W. and Lu, H. W. and Zheng, H. L.},
title = {Out-of-plane temperature-dependent resistivity studies on Tl-based superconductors},
journal = {Journal of Physics: Condensed Matter}}

@article{Honma2010,
title = {Universal scaling of the $c$-axis DC conductivity for underdoped high-temperature cuprate superconductors},
journal = {Solid State Communications},
volume = {150},
number = {47},
pages = {2314-2317},
year = {2010},
issn = {0038-1098},
doi = {10.1016/j.ssc.2010.10.003},
url = {https://www.sciencedirect.com/science/article/pii/S0038109810005922},
author = {T. Honma and P. H. Hor}}

@article{Herrmann1993,
title = {Single crystal studies of the pairing mechanism in Tl$_2$Ba$_2$CuO$_6$ superconductors},
journal = {Physica C: Superconductivity},
volume = {209},
number = {1},
pages = {199-202},
year = {1993},
issn = {0921-4534},
doi = {10.1016/0921-4534(93)90905-6},
url = {https://www.sciencedirect.com/science/article/pii/0921453493909056},
author = {A. M. Hermann and H. M. Duan and W. Kiehl and M. Paranthaman}
}

@article{Hunter2024,
    author = {Hunter, A. and Putzke, C. and Gaponenko, I. and Tamai, A. and Baumberger, F. and Moll, P. J. W.},
    title = {Controlling crystal cleavage in Focused Ion Beam shaped specimens for surface spectroscopy},
    journal = {Review of Scientific Instruments},
    volume = {95},
    number = {3},
    pages = {033905},
    year = {2024},
    month = {03},
    issn = {0034-6748},
    doi = {10.1063/5.0186480},
    url = {https://doi.org/10.1063/5.0186480}
}

@article{Diaz2022,
doi = {10.1088/1361-6463/ac357f},
url = {https://doi.org/10.1088/1361-6463/ac357f},
year = {2021},
month = {11},
publisher = {IOP Publishing},
volume = {55},
number = {8},
pages = {084001},
author = {Diaz, J. and Putzke, C. and Huang, X. and Estry, A. and Analytis, J. G. and Sabsovich, D. and Grushin, A. G. and Ilan, R. and Moll, P. J. W.},
title = {Bending strain in 3D topological semi-metals},
journal = {Journal of Physics D: Applied Physics}
}

@article{Analytis2007,
  title = {Angle-dependent magnetoresistance measurements in ${\mathrm{Tl}}_{2}{\mathrm{Ba}}_{2}\mathrm{Cu}{\mathrm{O}}_{6+\ensuremath{\delta}}$ and the need for anisotropic scattering},
  author = {Analytis, J. G. and Abdel-Jawad, M. and Balicas, L. and French, M. M. J. and Hussey, N. E.},
  journal = {Physical Review B},
  volume = {76},
  issue = {10},
  pages = {104523},
  numpages = {12},
  year = {2007},
  month = {09},
  publisher = {American Physical Society},
  doi = {10.1103/PhysRevB.76.104523},
  url = {https://link.aps.org/doi/10.1103/PhysRevB.76.104523}
}
